\documentclass[a4paper,11pt]{article}
\usepackage{jcappub}

\usepackage{mathrsfs}
\usepackage{graphicx}
\usepackage{amsmath,amssymb}
\usepackage{hyperref}
\usepackage{braket}
\usepackage{subfigure}
\usepackage{subfloat}
\usepackage{float}
\usepackage[usenames]{color}
\usepackage[normalem]{ulem}
\usepackage{booktabs}

\hypersetup{
    colorlinks=true,
    linkcolor=red,
    citecolor=blue,
}

\allowdisplaybreaks

\def\be{\begin{equation}}
\def\ee{\end{equation}}
\def\ba{\begin{eqnarray}}
\def\ea{\end{eqnarray}}

\definecolor{ForestGreen}{RGB}{36,179,0}

\title{Principal reconstructed modes of dark energy and gravity}

\author[1, 2]{Marco Raveri} 
\author[3, 4, 5]{Levon Pogosian}
\author[6, 7]{Matteo Martinelli}
\author[4]{Kazuya Koyama}
\author[8]{Alessandra Silvestri} 
\author[9, 10]{Gong-Bo Zhao}

\affiliation[1]{Department of Physics, University of Genova and INFN, Via Dodecaneso 33, 16146, Genova, Italy}
\affiliation[2]{Center for Particle Cosmology, Department of Physics and Astronomy, University of Pennsylvania, Philadelphia, PA 19104, USA}

\affiliation[3]{Department of Physics, Simon Fraser University, Burnaby, BC, V5A 1S6, Canada}
\affiliation[4]{Institute of Cosmology and Gravitation, University of Portsmouth, Portsmouth, PO1 3FX, UK}
\affiliation[5]{National Astronomy Observatories, Chinese Academy of Science, Beijing, 100101, P.R.China}

\affiliation[6]{INAF-Osservatorio Astronomico di Roma, Via Frascati 33, I-00078}
\affiliation[7]{Instituto de F\'isica Teorica UAM-CSIC, Campus de Cantoblanco, E-28049 Madrid, Spain}

\affiliation[8]{Institute Lorentz, Leiden University, PO Box 9506, Leiden 2300 RA, The Netherlands}

\affiliation[9]{National Astronomy Observatories, Chinese Academy of Science, Beijing, 100101, P.R.China}
\affiliation[10]{University of Chinese Academy of Sciences, Beijing, 100049, P.R.China}

\emailAdd{marco.raveri@unige.it}
\emailAdd{levon@sfu.ca}

\abstract{
Recently, in \cite{recon_letter}, we presented the first combined non-parametric reconstruction of the three time-dependent functions that capture departures from the standard cosmological model, $\Lambda$CDM, in the expansion history and gravitational effects on matter and light from the currently available combination of the background and large scale structure data. The reconstruction was performed with and without a theory-informed prior, built on the general Horndeski class of scalar-tensor theories, that correlates the three functions. 
In this work, we perform a decomposition of the prior and posterior covariances of the three functions to determine the structure of the modes that are constrained by the data relative to the Horndeski prior. 
We find that the combination of all data can constrain 15 combined eigenmodes of the three functions with respect to the prior. 
We examine and interpret their features in view of the well-known tensions between datasets within the $\Lambda$CDM model. We also assess the bias introduced by the simplistic parameterizations commonly used in the literature for constraining deviations from GR on cosmological scales.
}

\begin{document}

\maketitle
\flushbottom

\section{Introduction}
Rapidly improving data from cosmological surveys is opening new opportunities for testing the pillars of the $\Lambda$ Cold Dark Matter ($\Lambda$CDM) model. Along with probing the nature of dark matter and dark energy (DE), it is becoming possible to examine the foundational principles of General Relativity (GR), such as the universal geometric nature of gravity and the precise way in which matter distorts spacetime. Since the discovery of cosmic acceleration~\cite{Riess:1998cb,Perlmutter:1998np}, significant effort went into constraining the dynamics of DE, primarily by looking for deviations of its equation of state from $w_\Lambda=-1$. The past decade and a half also witnessed the emergence and maturing of the field of cosmological tests of GR, which led to identifying broad classes of potentially interesting modified gravity (MG) theories (see~\cite{Silvestri:2009hh,Clifton:2011jh,Joyce:2014kja,Koyama:2015vza,Ishak:2018his} for reviews) and developing phenomenological frameworks for non-model-specific tests~\cite{Amendola:2007rr,Bertschinger:2008zb,Pogosian:2010tj,Gubitosi:2012hu,Bloomfield:2012ff,Gleyzes:2013ooa,Gleyzes:2014rba,Bellini:2014fua,Amendola:2012ky,Lagos:2016wyv} along with their numerical implementations~\cite{Zhao:2008bn,Hojjati:2011ix,Hu:2013twa,Raveri:2014cka,Zumalacarregui:2016pph}. Testing gravity and the physics of DE is one of the primary science goals of the ongoing and upcoming surveys, such as DESI~\cite{DESI}, Euclid~\cite{euclid} and Vera Rubin Observatory~\cite{lsst}, which will take these tests to qualitatively higher levels \cite{Alam:2020jdv,EuclidTheoryWorkingGroup:2012gxx,Ishak:2019aay,Martinelli:2021hir}.

Until recently, the majority of phenomenological tests of DE dynamics and departures from GR were conducted independently from each other. Namely, GR would be assumed when constraining the evolution of $w$ with redshift $z$, or $w=-1$ would be assumed when constraining the MG effects in the growth of structure, parameterized, {\it e.g.}, by phenomenological functions $\mu(k,z)$ and $\Sigma(k,z)$~\cite{Garcia-Quintero:2020bac}. In addition, in most cases, fixed simple parametric forms were used for $w(z)$, $\mu(z)$ and $\Sigma(z)$ or their equivalents. Such a simplification may be justified when the constraining power of the data is limited -- {\it e.g.} measurements of the cosmic microwave background (CMB) temperature and polarization anisotropies only constrain a weighted average of $w(z)$, hence it makes sense to fit a constant $w$ to CMB alone. However, measurements of the baryon acoustic oscillations (BAO), supernovae (SN) magnitudes, as well as the galaxy counts and galaxy shear surveys, offer measures of the background expansion and the growth of large scale structure at multiple redshifts. Using simple parameterizations when analyzing these datasets is likely to bias the outcome and result in a loss of potentially important information. In addition, in any specific MG theory, the dynamics of the effective DE, which impacts the background expansion, is correlated with the changes in the growth of perturbations. Thus, rather than assuming that only one of the two is modified, it makes more physical sense to vary them simultaneously when performing fits to the data. As was demonstrated in \cite{recon_letter}, current datasets already allow us to simultaneously reconstruct the effective DE density and the two modified growth functions using flexible non-parametric forms.

There are several approaches to the non-parametric reconstruction of cosmological functions such as $w(z)$. Popular methods include binning the functions at several redshifts, using Gaussian Processes (GP)~\cite{Shafieloo:2012ht,Seikel:2012uu,Gerardi:2019obr}, and the correlated prior approach~\cite{Crittenden:2011aa,Zhao:2012aw,Zhao:2017cud,Wang:2018fng}. A simple binning, with the function assumed to be constant and independent in each bin, or with a smooth interpolation between the redshift nodes, makes the results dependent on an unphysical implicit smoothness prior. Also, one is typically restricted in the number of bins they can use by the MCMC convergence times. Using a small number of bins might, in turn, bias the reconstruction. The GP method is not restricted in the number of bins, introducing instead a Gaussian prior that correlates the function at neighbouring redshifts. However, the choice of the GP prior is essentially phenomenological (without any relation to a physical theory) and the parameters of the prior are typically marginalized over, thus obscuring the Bayesian interpretation of the resultant reconstruction. The correlated prior approach also introduces a correlation between the neighboring redshifts but, unlike the GP method, it uses a fixed prior covariance matrix which is meant to be derived from theory. Having an explicit prior makes it possible to clearly state how much the data improves on the prior, and to compute the Bayesian evidence that can be compared to that of the $\Lambda$CDM model. 

In this work, we examine the joint reconstruction of the redshift dependence of the effective DE fractional energy density $\Omega_X(z)$ and the phenomenological functions $\mu$ and $\Sigma$, which parameterize possible modifications of the Poisson equation relating the matter density contrast to the Newtonian and the lensing potentials, respectively, performed from the combination of the current CMB, BAO, Redshift Space Distortion (RSD), SN, galaxy counts and galaxy weak lensing data in \cite{recon_letter}. Since the phenomenological parameterization by $\mu$ and $\Sigma$ is valid only for linear perturbations, such an analysis is restricted to linear scales. The reconstruction was performed with and without using a prior covariance of the three functions derived from the Hordenski class of theories~\cite{Espejo:2018hxa}. It was shown that the theoretical prior has a significant smoothing effect on the allowed variation of these functions with redshift. 
We perform a decomposition in Covariant Principal Components~\cite{Dacunha:2021uig} of the prior and posterior covariances of the three functions to determine the number of modes that are constrained by the data relative to the Horndeski prior, and find that current data can constraint 15 independent degrees of freedom (DOF) (combined eigenmodes) of the three functions relative to the prior. This means one can learn much more from the current data than one would if using simple {\it ad hoc} parametric forms. We also asses the bias introduced by using some of the commonly used parametric forms.

This paper is organized as follows. 
In Sec.~\ref{sec:recon} we briefly review the phenomenological functions $\Omega_X$, $\mu$ and $\Sigma$ and the reconstruction performed in \cite{recon_letter}. 
We then present the CPCA decomposition and discuss the significance of features in Sec.~\ref{sec:eigen}. 
The analysis of the bias introduced by simplistic parametric forms is presented in Sec.~\ref{Sec:Bias}, followed by a concluding summary in Sec.~\ref{Sec:Summary}.

\section{Reconstructed gravity} 
\label{sec:recon}

\subsection{Model}
\label{sec:model}

The reconstruction of \cite{recon_letter} concerned the background expansions and linear scalar perturbations to the Friedmann-Lemaitre-Roberston-Walker (FLRW) metric. Working in the Newtonian gauge, the line-element reads:
\be\label{line_element}
ds^2 = -(1+2\psi)dt^2 + a^2(1-2\phi) d{\bf x}^2 \ ,
\ee
where $a$ is the scale factor, $t$ is the cosmic time, and $\psi$ and $\phi$ are the scalar perturbations to the metric. The dynamics of the expansion is set by the Friedmann equation,
\be\label{Friedmann}
\frac{H^2}{H_0^2}=\Omega_r(1+z)^4+\Omega_m(1+z)^3+\Omega_{\rm X}(z)\,,
\ee
where $H=d\ln{a}/dt$ is the Hubble parameter, $H_0$ is its current value, $z=1/a-1$ is the redshift, $\Omega_r$ and $\Omega_m$ are the fractional energy densities of radiation and  matter. The time evolution of the fractional energy density of dark energy is described by $\Omega_X(z)$ with $\Omega_{\rm DE} \equiv \Omega_{\rm X}(z=0)$, the fractional energy density of dark energy today. 
We assume spatial flatness, hence $\Omega_r+\Omega_m+\Omega_{\rm DE}=1$. 
It is worth noting that the DE component is defined quite broadly through this equation, to capture not only a dynamical DE field, but also, {\it e.g.}, modifications to gravity or non-minimal interactions with matter. In other words, $\Omega_X(z)$ represents an {\it effective DE fluid} that encodes all contributions to the Friedmann equation different from radiation and minimally coupled matter, with $\Omega_X(z)=\Omega_{\rm DE}$ corresponding to the cosmological constant $\Lambda$. As emphasized in~\cite{Wang:2018fng}, using the DE equation of state to describe such an effective fluid could potentially bias the reconstruction because it prohibits the effective DE density from changing its sign, which is not uncommon in theories with new interactions. This is the reason for choosing to work with $\Omega_X(z)$ instead of $w(z)$. 

The linearly perturbed Einstein equations provide a set of equations relating metric perturbations to the perturbations in the energy-momentum tensor of the matter fields. As shown in~\cite{Amendola:2007rr,Bertschinger:2008zb,Pogosian:2010tj}, the phenomenology of linear perturbations in many extensions of $\Lambda$CDM can be effectively captured by introducing two functions of time and scale, defined through the Poisson equations in Fourier space for the Newtonian potential $\psi$ and the lensing (Weyl) potential $\phi+\psi$, as
\ba
&& k^2 \psi = - 4 \pi G  \mu(a, k) a^2 \left[ \rho \Delta + 3 (\rho+P) \sigma \right] \ ,\\
\label{poisson_mg}
&& k^2 (\phi + \psi) = - 4 \pi G \Sigma(a,k) a^2  \left[ 2\rho \Delta + 3(\rho+P) \sigma \right]  \ ,
\label{shear_mg}
\ea
where $k$ is the Fourier number, $G$ is the gravitational constant, $\rho$ is the background matter density, and $\Delta$ is the comoving density contrast, $\Delta \equiv \delta + 3aHv/k$, where $\delta \equiv \delta \rho/\rho$ is the density contrast in the Newtonian gauge and $v$ is the irrotational component of the peculiar velocity. The anisotropic stress $\sigma$ due to relativistic components is included for consistency but is negligible during matter and dark energy dominated epochs. Since $\Sigma$ directly controls the Weyl potential, it is best constrained by weak lensing (WL) measurements. On the other hand, $\mu$ sets the Newtonian potential, which determines the peculiar velocities of galaxies. Thus, combining the WL data with RSD allows us to effectively break the degeneracy between the two functions~\cite{Amendola:2007rr,Pogosian:2010tj,Song:2010fg}.

\subsection{Reconstruction} 
\label{Sec:Methods}

\begin{figure*}[tbph!]
\centering
\includegraphics[width=0.48\columnwidth]{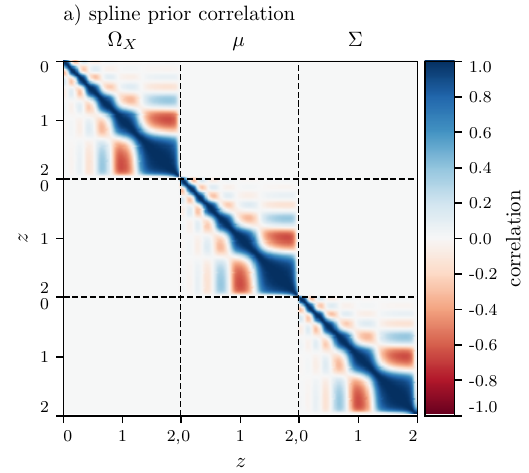}
\includegraphics[width=0.48\columnwidth]{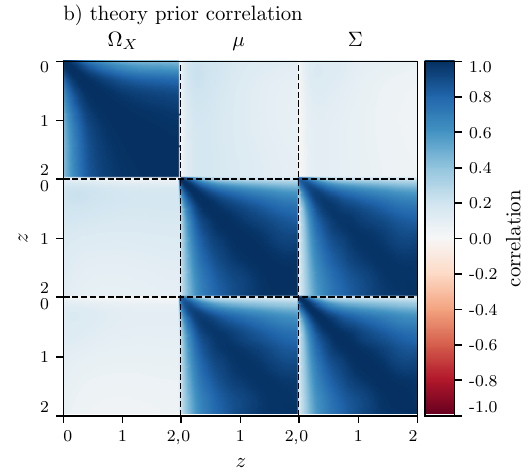}
\includegraphics[width=0.48\columnwidth]{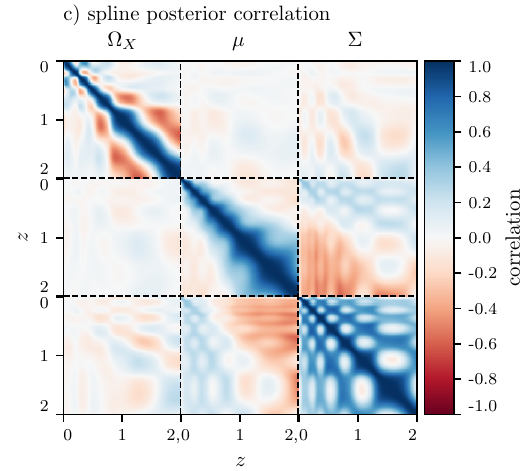}
\includegraphics[width=0.48\columnwidth]{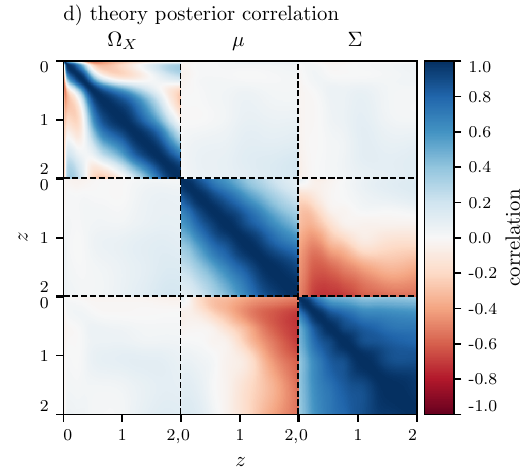}
\includegraphics[width=0.48\columnwidth]{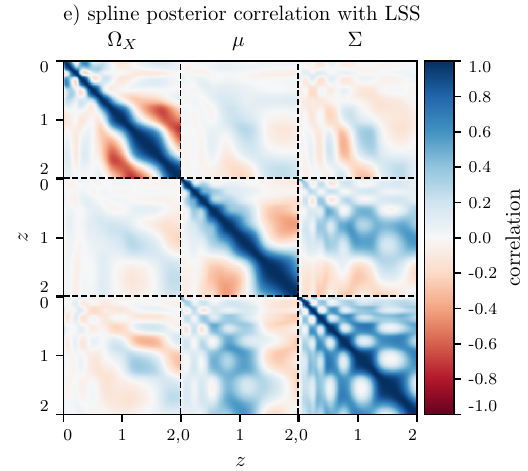}
\includegraphics[width=0.48\columnwidth]{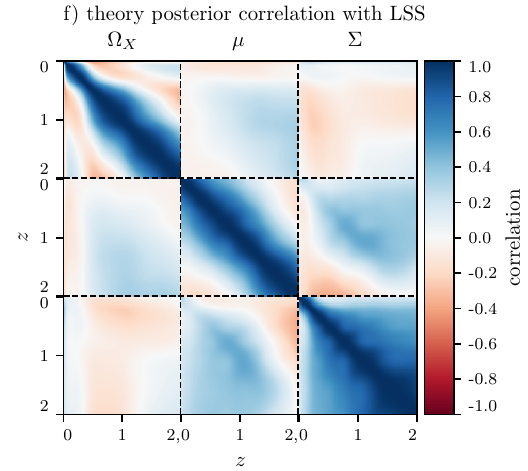}
\caption{ \label{fig:prior_correlations}
a) The implicit correlation prior, as a function of redshift, induced by the cubic spline on the functions $\Omega_X$, $\mu$ and $\Sigma$. All functions have the same implicit prior independently of each other.
b) The Horndeski prior correlation. The correlation between each function is much stronger than that introduced by the cubic spline. In the Horndeski prior we also see a strong correlation between $\mu$ and $\Sigma$. 
c) The correlation obtained from our ``Baseline'' data posterior covariance of the nodes, {\it i.e.} that determined by the data and the implicit prior correlation in Panel (a). 
d) The correlation corresponding to the posterior covariance derived from the Baseline data with the help of the Horndeski prior in Panel (b).
e)/f) same as Panels (c)/(d) for the ``Baseline+LSS'' data combination. 
}
\end{figure*}

A joint reconstruction of $\Omega_X(a)$, $\mu(a)$ and $\Sigma(a)$ from the currently available data was recently performed in \cite{recon_letter}. Here we briefly review their method. 
The three functions were represented by their values at $11$ discrete values (nodes) of $a$, with a cubic spline used to interpolate between them. From the $11$ nodes, $10$ values are distributed uniformly in the interval $a\in[1,0.25]$ (corresponding to $z\in [0,3]$) with another one at $a=0.2$ ($z=4$). The functions were made to approach their $\Lambda$CDM values at higher redshifts, although studying earlier times deviations from GR is generally possible within the same framework~\cite{Lin:2018nxe}. The cubic spline introduces correlations between the nodes shown in Panel (a) of Fig.~\ref{fig:prior_correlations}.

In addition to performing the reconstruction of $\Omega_X(a)$, $\mu(a)$ and $\Sigma(a)$ from the data alone, we used the method of~\cite{Crittenden:2011aa,Crittenden:2005wj} to add the Horndeski prior that correlates the nodes $\{\Omega_{Xi},\mu_i,\Sigma_i \} \equiv {\bf f}$. It is introduced as a Gaussian prior
\be
{\cal P}_{\rm prior} \propto \exp [-({\bf f}-{\bf f}_{\rm fid}) {\cal C}^{-1} ({\bf f}-{\bf f}_{\rm fid})^T] \ ,
\ee
where ${\cal C}$ is the correlation matrix derived from the joint covariance of the three functions obtained in~\cite{Espejo:2018hxa}. The theory prior acts much like a Wiener filter, discouraging (but not completely prohibiting) abrupt variations of the functions. Panel (b) of Fig.~\ref{fig:prior_correlations} shows the Horndeski correlation prior used in this work. One can clearly see that the correlation ``length'' is much longer than that of the implicit prior due to the cubic spline shown in Panel (a). This ensures that the prior aided reconstruction is independent of the binning scheme. 

The reconstruction was performed using an appropriately modified version of {\tt MGCosmoMC}\footnote{\url{https://github.com/sfu-cosmo/MGCosmoMC}}~\cite{Zhao:2008bn,Hojjati:2011ix,Zucca:2019xhg}, based on {\tt CosmoMC}\footnote{\url{http://cosmologist.info/cosmomc/}}~\cite{Lewis:2002ah}, to sample the parameter space, which, in addition to the node parameters $\Omega_{Xi}$, $\mu_i$, $\Sigma_i$ introduced earlier, includes the usual cosmological parameters: $\Omega_bh^2$, $\Omega_ch^2$, $\theta_\star$, $\tau$, $A_s$, $n_s$, $\mathcal{N}$, where $\Omega_{b}h^2$ and $\Omega_{c}h^2$ are the physical densities of baryons and CDM, $\theta_\star$ is the angular size of the sound horizon at the decoupling epoch, $\tau$ is the reionization optical depth, $A_s$ and $n_s$ are the amplitude and the spectral index of primordial fluctuations, and $\mathcal{N}$ collectively denotes the nuisance parameters that appear in various data likelihoods. 

The following datasets were considered:
\begin{itemize}
\item ``Planck'': the 2018 release of the CMB temperature, polarization and the reconstructed CMB weak lensing spectra~\cite{Aghanim:2019ame};
\item ``BAO'': the eBOSS  DR16 BAO compilation from~\cite{Alam:2020sor} that includes measurements at multiple redshifts from the samples of Luminous Red Galaxies (LRGs), Emission Line Galaxies (ELGs), clustering quasars (QSOs), and the Lyman-$\alpha$ forest~\cite{Zhao:2020tis,Wang:2020tje,Hou:2020rse,duMasdesBourboux:2020pck}, along with the SDSS DR7 MGS~\cite{Ross:2014qpa} data. We also add the BAO measurement from 6dF~\cite{Beutler:2011hx}. 
This compilation covers the BAO measurements at $0.07 < z < 3.5$. Note that the BAO data considered here are the ``tomographic''  version of the DR12 BOSS BAO at  $0.20 < z < 0.75$ \cite{BOSS:2016lpe} (not the ``consensus'' version using effective redshifts presented in \cite{Alam:2020sor}).
\item ``SN'': the Pantheon SN sample at $0.01 < z < 2.3$~\cite{Scolnic:2017caz};
\item ``RSD'': the eBOSS joint measurement of BAO and RSD for LRGs, ELGs and QSOs~\cite{Bautista:2020ahg,deMattia:2020fkb, Hou:2020rse,Neveux:2020voa}, using it instead of the eBOSS BAO-only measurement. For LRGs, it combines eBOSS LRGs and BOSS CMASS galaxies spanning the redshift range $0.6<z<1$, at an effective redshift of $z_{\rm eff}=0.698$. QSOs cover $0.8<z<2.2$ with an effective redshift of $z_{\rm eff}=1.48$, while ELGs cover $0.6<z<1.1$ with an effective redshift of $z_{\rm eff}=0.845$. In addition, we add BAO-only measurements from 6dF and MGS. 
\item ``DES'': the Dark Energy Survey Year 1 measurements of the angular two-point correlation functions of galaxy clustering, cosmic shear and galaxy-galaxy lensing with source galaxies at $0.2< z < 1.3$~\cite{Abbott:2017wau}; since our formalism has no nonlinear prescription for structure formation, the angular separations probing the nonlinear scales were removed using the ``aggressive'' cut option of {\tt MGCAMB} described in~\cite{Zucca:2019xhg}, which uses the method introduced in~\cite{Planck:2015bue,Abbott:2017wau}.
\end{itemize}
The baseline dataset combination (labelled ``Baseline'') includes Planck, BAO and SN. In addition, we also considered Baseline+RSD+DES. Note that, when RSD is included in the combination, the BAO data do not coincide with the one used in Baseline for the eBOSS LRGs BAO measurement, as we replaced it with the joint RSD-BAO measurement. For brevity, RSD+DES is referred to as simply ``LSS''.
We do not include the SH0ES determination of the intrinsic SN type Ia brightness magnitude as obtained by~\cite{Riess:2020fzl} in our analysis. The impact of the inclusion of this measurement and its implication for the Hubble constant tension can be found in \cite{recon_letter}.
\begin{figure*}[tpbh!]
\centering
\includegraphics[width=0.99\columnwidth]{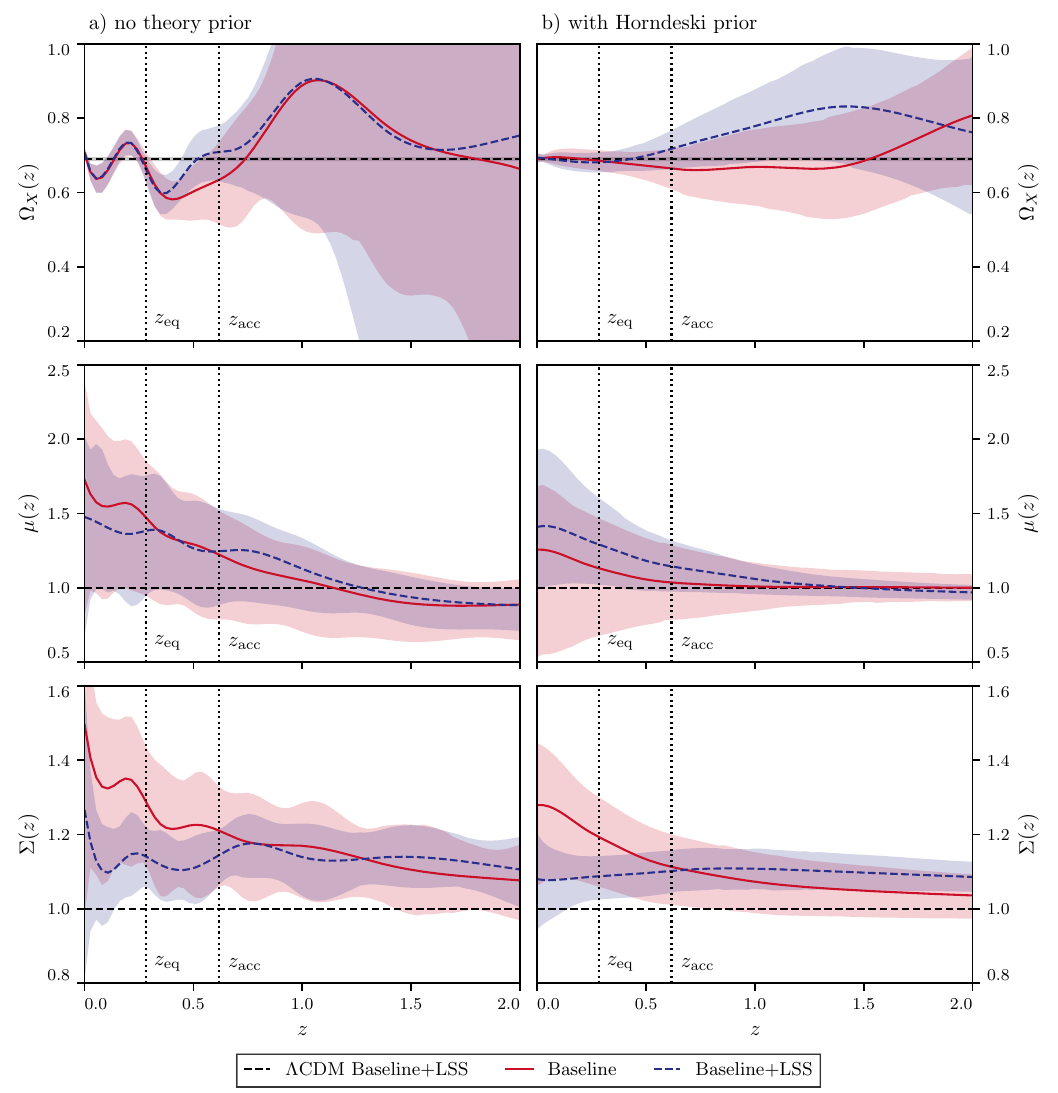}
\caption{Reconstruction of $\Omega_X(z)$ (top panels), $\mu(z)$ (middle panels) and $\Sigma(z)$ (bottom panels) from the Baseline (red lines) and Baseline+LSS (blue lines) data, without (left panels) and with the Horndeski correlation prior (right panels). The shaded regions show the $68\%$ confidence levels. The two vertical lines show the redshifts of equality between the matter and DE densities, $z_{\rm eq}$, and the beginning of cosmic acceleration, $z_{\rm acc}$, in the best fit $\Lambda$CDM model.}
\label{fig:b_vs_brd}
\end{figure*}

Fig.~\ref{fig:b_vs_brd} shows the functions $\Omega_X(z)$, $\mu(z)$ and $\Sigma(z)$ reconstructed from the Baseline and Baseline+LSS data combinations with and without the Horndeski correlation prior. One can see that the correlation prior smooths out the oscillations seen in the data-only reconstructions of all three functions. These oscillations are dependent on the number of nodes and affected by the implicit cubic-spline-induced correlation shown in Fig.~\ref{fig:prior_correlations}(a). 
The impact of the implicit prior is apparent for the data-only case shown in Fig.~\ref{fig:prior_correlations}(c), where the pattern of the rings evidently has the same frequency as the features in Fig.~\ref{fig:prior_correlations}(a). One can also see from Fig.~\ref{fig:prior_correlations}(c) that data (the Baseline dataset in this case) introduces correlations among the three functions, with $\Sigma$ being more strongly correlated with $\Omega_X$ than $\mu$. The data also introduces correlations between $\mu$ and $\Sigma$.  The artifacts of the implicit prior are not present when data is combined with the Horndeski prior, as shown in Fig.~\ref{fig:prior_correlations}(d). The theory prior suppresses correlations introduced by the cubic spline, while retaining the correlation introduced by the data. This shows the important role played by the theory prior in the reconstruction, as it prevents an overfitting of the data by favouring the reconstruction of only those features that are consistent with the theory. Fig.~\ref{fig:prior_correlations}(e) and Fig.~\ref{fig:prior_correlations}(f) show the effect of including the LSS measurement. It mainly affects the cross-correlations among the three functions as well as the correlation of $\mu$.

\section{Significance of the detection and the CPCA decomposition}\label{sec:eigen}

To gain further insight into the features and the number of DOF of the three functions constrained by the data, we use the Covariant Principal Components Analysis (CPCA), as discussed in~\cite{Dacunha:2021uig}.
This decomposes the posterior covariance in units of the prior covariance to ensure the independence of results from the specific parametrization that is used.

More specifically we decompose the prior $\mathcal{C}_\Pi$ and posterior $\mathcal{C}_p$ covariances as
\begin{align}
\mathcal{C}_{\Pi} \Psi = \mathcal{C}_p \Psi \Lambda ,
\end{align}
where the matrix $\Psi$ has as columns the CPC eigenmodes of the posterior with respect to the prior, and the matrix $\Lambda$ is diagonal and quantifies the improvement of the posterior with respect to the prior.
While the CPC modes are not orthonormal in the Euclidean sense, they are orthogonal in the metrics induced by the prior and posterior covariances:
\begin{align}
\Psi^T \mathcal{C}_\Pi \Psi =& \Lambda \nonumber \\
\Psi^T \mathcal{C}_p \Psi =& I ,
\end{align}
so that parameters projected along the CPC modes are statistically independent for both the prior and the posterior.

\begin{table}[!ht]
\setlength{\tabcolsep}{8pt}
\centering
\begin{tabular}{@{}cccccc@{}}
\toprule
                           & $\Omega_X$ & $\mu$ & $\Sigma$ & Combined \\
\toprule              
Baseline $N_{\rm eff}$     & 5.6        & 1.7   & 7.6      & 14.9 \\
Baseline+LSS $N_{\rm eff}$ & 5.6        & 2.0   & 8.3      & 15.3 \\                                                                   
\midrule
Baseline $T$               & 1834.4     & 22.7  & 141.7    & 2603.6 \\
Baseline+LSS $T$           & 1816.8     & 40.3  & 170.3    & 2755.6 \\    
\bottomrule
\end{tabular}
\caption{\label{tab:pca_total_snr} 
The number of well-constrained eigenmodes of $\Lambda$, $N_{\rm eff}$, for each function and for the three combined, along with the trace, $T = {\rm Tr}(\Lambda)$, that quantifies the net constraining power of the data}
\end{table}

\begin{figure}[tbph!]
\centering
\includegraphics[width=0.5\columnwidth]{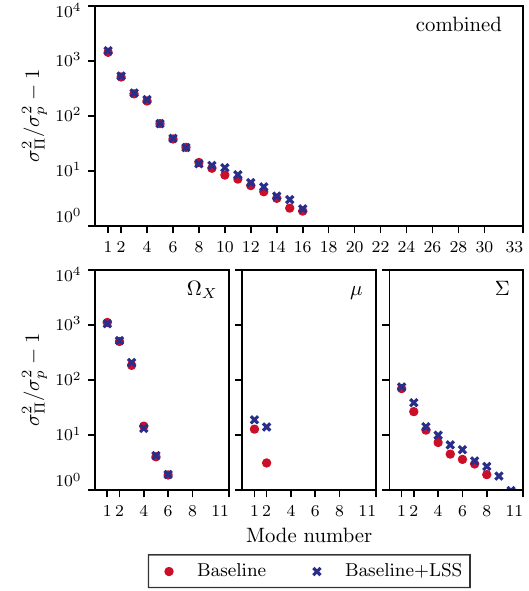}
\caption{\label{fig:evalues} 
The eigenvalues of the improvement matrix $\Lambda$ computed from the posterior and the prior covariances of the three functions jointly and individually.
}
\end{figure}

\begin{figure*}[tbph!]
\centering
\includegraphics[width=0.99\textwidth]{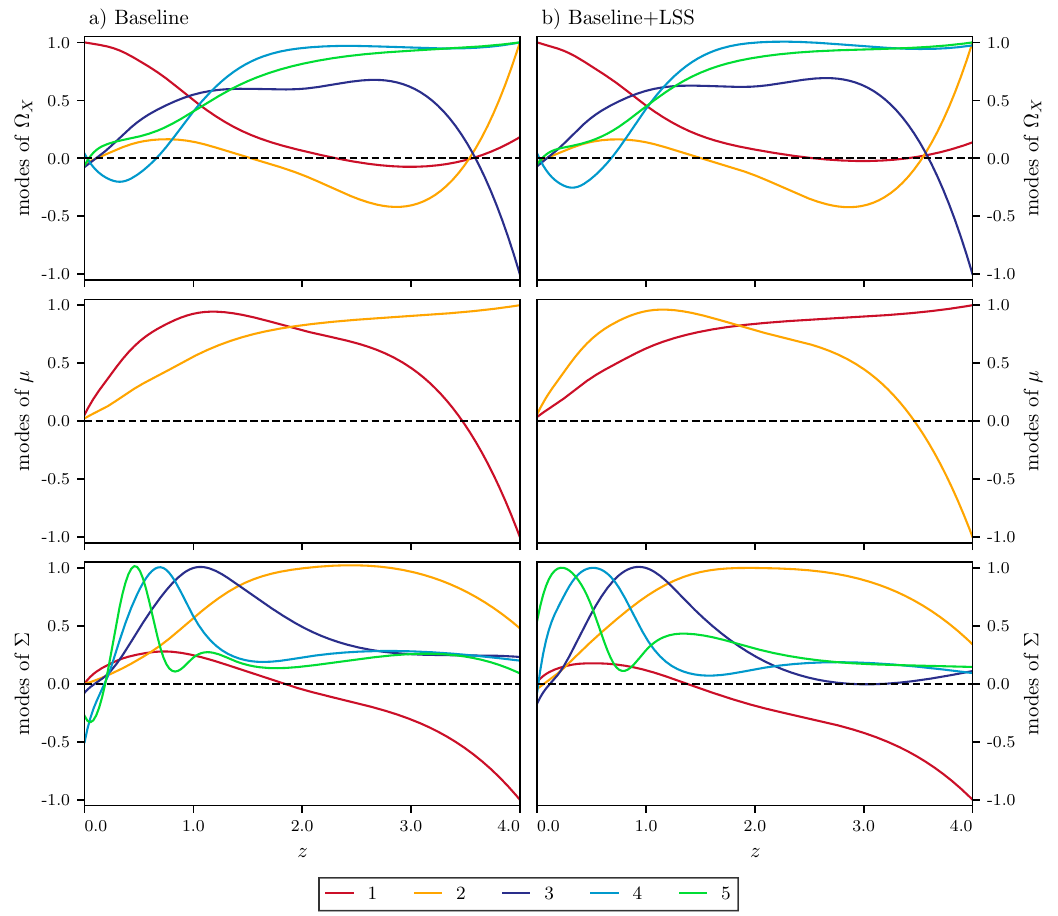}
\caption{\label{fig:emodes} 
The shapes of the best constrained individual eigenmodes of $\mu$, $\Sigma$ and $\gamma$ plotted vs redshift for the Baseline (left) and Baseline+LSS (right). The amplitude and the overall sign of modes are arbitrary and are rescaled in the figure so that the maximum of each mode is one. The eigenmodes can be interpreted as the window functions representing sensitivity of the data to the redshift evolution of the three functions.
}
\end{figure*}

\begin{figure}[tbph!]
\centering
\includegraphics[width=0.5\columnwidth]{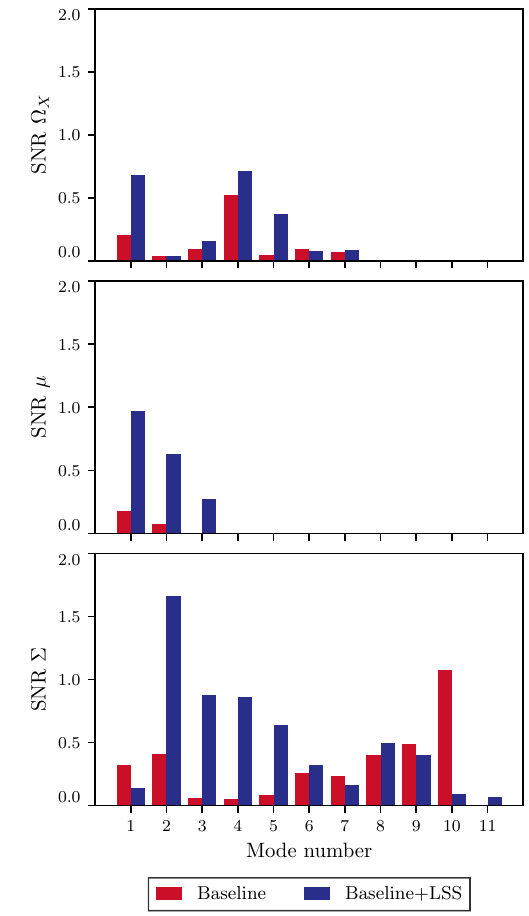}
\caption{\label{fig:eigen_snr}
The SNR of deviations from $\Lambda$CDM for each individual constrained mode, for Baseline and Baseline+LSS. 
}
\end{figure}

The posterior Fisher matrix can also be decomposed into the same CPC modes:
\begin{align}
\mathcal{C}_p^{-1} =& \Psi \Psi^T ,
\end{align}
so that we can easily compute the Signal to Noise Ratio (SNR) as a sum of the SNR in each of the independent CPC modes:
\begin{align} \label{eq:theta_snr}
{\rm SNR}^2 \equiv (\theta-\theta_0)^T \mathcal{C}_p^{-1} (\theta-\theta_0)= \sum_a (\Psi_a^T (\theta-\theta_0))^2
\end{align}
where $\theta_0$ is the reference vector for the SNR calculation and the index $a$ spans the CPC modes.

Following~\cite{Raveri:2018wln}, we can also define the number of parameters in our reconstruction that are constrained relative to the prior as
\begin{align}
N_{\rm eff} = N - {\rm Tr}(\mathcal{C}_\Pi^{-1} \mathcal{C}_p) ,
\end{align}
where $N$ is the number of nominal parameters. This is a coarse measure of the constraining ability of a measurement, because it only counts how many parameters are appreciably constrained, rather that quantifying how well they are constrained. Along with this, we introduce a quantitative measure of the constraining power, $T$, defined as the trace of the ``improvement'' matrix $\Lambda$, 
\begin{align}
T = {\rm Tr}(\Lambda) .
\end{align}

Fig.~\ref{fig:evalues} shows the eigenvalues of $\Lambda$ ordered from highest to lowest, thus corresponding to the best-to-worst constrained eigenmodes $\Psi$. They can be written as $\lambda_i = \sigma^2_{\Pi,i}/\sigma^2_{p,i}-1$, where $\sigma^2_{\Pi,i}$ and $\sigma^2_{p,i}$ are the eigenvalues of the prior and the posterior covariances, so that a mode can be considered ``constrained'' relative to the prior when $\lambda_i \gtrsim 1$.
We show the eigenvalues of the ``combined'' modes, corresponding to the joint covariance of all three functions, as well as those for the individual functions, after marginalizing over the other two. The former tells us how many independent DOF (roughly) quantifying departures from $\Lambda$CDM can be measured without asking what function they correspond to. The latter quantify the ability of data to constrain the specific functions. One can see that the number of constrained combined modes is quite large, around 15, and that the top highest eigenvalues are the same as those for $\Omega_X$ that probes the background expansion. Also, it is clear that $\Sigma$ is much better constrained than $\mu$. 

Interestingly, the number of constrained modes does not change appreciably with the inclusion of the LSS data. This is, in part, because our $N_{\rm eff}$ is only a coarse measurement of improvement. Also, since we have cut the LSS data to exclude nonlinear scales, the CMB and LSS are probing a similar range of scales. Hence, most of the modes that {\emph can} be constrained on linear scales are already constrained, to some extent, by the Baseline data. The addition of the LSS data, however, makes a notable difference in how well the individual modes of $\mu$ and $\Sigma$ are constrained. As one can see from Table~\ref{tab:pca_total_snr}, the trace $T$ is increased by a factor of $\sim$2 for $\mu$ and by 20\% for $\Sigma$. This illustrates that combining the RSD data and WL helps to break the degeneracy between $\mu$ and $\Sigma$.

Further insight can be gained by considering the shapes of the best constrained eigenmodes $\Psi$ when plotted as functions of redshift. They can be interpreted as the window functions representing sensitivity of the data to the redshift evolution of $\mu$, $\Sigma$ and $\gamma$. As one can see from Fig.~\ref{fig:emodes}, the modes derived from Baseline and Baseline+LSS appear rather similar. For $\Omega_X$, in particular, the change is difficult to detect by eye. This is because the LSS constraint on the expansion history is much weaker than that of Baseline. The redshifts covered by the top three modes of $\Omega_X$, in the order from best to worst constrained, are low-$z$, high-$z$, and in between.

The two constrained eigenmodes of $\mu$ can be identified with the overall growth of structure and the ISW effect. In both cases, the impact of $\mu$ is an integrated effect, {\it i.e.} via the change of the gravitational coupling in the differential equation that determines the growth of density perturbations. Hence, both modes have a broad support in redshift. Interestingly, the addition of LSS flips the two modes -- the ``ISW'' mode, best constrained by Baseline, becomes the second best, since LSS includes additional measurements of WL and RSD. 

The top two modes of $\Sigma$ mirror those of $\mu$, but with different pivot points, since the impact of $\Sigma$ on the Weyl potential is direct, not integrated like in the case of $\mu$. The higher order modes match quite well the lensing kernels that contribute to the lensing of the CMB temperature and polarization anisotropies (see Fig.~11 of \cite{Planck:2018lbu}). With the addition of LSS, the CMB kernels that correspond to the large scale CMB lensing, {\it i.e.} occurring at lower redshifts, become mixed with the galaxy lensing kernels of DES, but the first few best modes are largely unchanged in shape. As mentioned earlier, the ability to constrain the modes of $\mu$ and $\Sigma$ increases appreciably with the addition of LSS.

Finally, Fig.~\ref{fig:eigen_snr} shows the SNR of deviations from $\Lambda$CDM for each mode.

There are several well-known tensions in $\Lambda$CDM. The first is the disagreement in the galaxy clustering amplitude, quantified by the parameter $S_8$, predicted by the best fit to CMB and that measured by galaxy weak lensing surveys. In addition, the CMB temperature anisotropy measured by Planck appears to be more affected by weak gravitational lensing than expected in $\Lambda$CDM~\cite{Planck:2018vyg}, known as the $A_L$ anomaly. Finally, the temperature (TT) power spectrum  at low multipoles is lower than the $\Lambda$CDM prediction.

In the case of $\Omega_X$, for both Baseline and Baseline+LSS, the most anomalous modes are the first and the fifth, which have support at low and high redshifts. 
These are representative of the low TT power at low multipoles and the $A_L$ anomaly, respectively. The most anomalous mode of $\mu$ is also the one that is best constrained, corresponding to the net growth of the large scale structure and, therefore, most affected by the $S_8$ tension. Essentially the same mode is also the most anomalous for $\Sigma$, where it is the the second best constrained. In all cases, the significance of detection is increased with the addition of the LSS.

\section{Bias introduced by simple parametrizations} \label{Sec:Bias}

It is interesting to check how well our reconstructed functions can be fit by simple parameterizations. For $\mu$ and $\Sigma$, we consider the following forms:
\begin{enumerate}
\item constant $\mu$ and $\Sigma$, although this parametrization is not commonly used; 
\item the ``DE fraction'' parametrization used by DES~\cite{DES:2018ufa} and Planck~\cite{Planck:2015bue}, where the time dependence of $\mu$ and $\Sigma$ is determined to the fraction of the total energy density in DE, i.e. 
\be
\mu,\Sigma = 1+\alpha_{\mu,\Sigma} \rho_{\rm DE}(a)/\rho_{\rm tot}(a) \ ; \nonumber
\ee
\item the ``linear model'' used by Planck~\cite{Planck:2015bue}, where
\be
\mu,\Sigma = 1+\alpha_{\mu,\Sigma} +\beta_{\mu,\Sigma} (1-a)  \ ; \nonumber
\nonumber
\ee
\end{enumerate}
For the effective DE evolution, we consider two commonly used parametrization DE equation of state:
\begin{enumerate}
\item constant $w$;
\item the Chevallier-Polarski-Linder (CPL) parametrization~\cite{Chevallier:2000qy,Linder:2002et}
\be
w(a) = w_0+w_a (1-a) . \nonumber
\ee
\end{enumerate}
We then take the reconstruction MCMC chains and, sample by sample, project them on the above parameterizations.

\begin{table}[!ht]
\setlength{\tabcolsep}{12pt}
\centering
\begin{tabular}{@{}cccccc@{}}
\toprule
			   	    & SNR$^2$ & Bias \\
\toprule
no theory prior      & & \\
\midrule
$w$ constant         & 4.7 & 4.6 \\ 
$w = w_0 + w_a(1-a)$ & 4.7 & 4.6 \\
$\mu$ constant       & 3.1 & 2.7 \\
$\mu = 1 +\alpha \Omega_{\rm DE}$ & 3.1 & 2.7 \\
$\mu = 1 +\alpha +\beta(1-a)$ & 3.1 & 0.7 \\
$\Sigma$ constant    & 6.2 & 4.3 \\
$\Sigma = 1 +\alpha \Omega_{\rm DE}$ & 6.2 & 4.4 \\
$\Sigma = 1 +\alpha +\beta(1-a)$ & 6.2 & 3.0 \\
\toprule
with Horndeski prior & & \\
\midrule 
$w$ constant         & 1.0 & 1.0 \\ 
$w = w_0 + w_a(1-a)$ & 1.0 & 1.0 \\
$\mu$ constant       & 2.5 & 1.9 \\
$\mu = 1 +\alpha \Omega_{\rm DE}$ & 2.5 & 2.5 \\
$\mu = 1 +\alpha +\beta(1-a)$ & 2.5 & 0.4 \\
$\Sigma$ constant    & 5.3 & 2.4 \\
$\Sigma = 1 +\alpha \Omega_{\rm DE}$ & 5.3 & 4.3 \\
$\Sigma = 1 +\alpha +\beta(1-a)$ & 5.3 & 1.8 \\
\bottomrule
\end{tabular}
\caption{\label{tab:bias}
Bias introduced by using a simple parameterization for the Baseline+LSS dataset, as defined by Eq.~\ref{eq:bias} and the SNR$^2$ for the corresponding reconstruction, which is the upper bound of Bias. 
The closer the two values are, the worse is the ability of the parameterization to represent the function.
For $w$, to avoid singularities caused by negative effective DE densities, the comparison was restricted to $z\in [0,1]$ and $[0,2]$ for the cases without and with the Horndeski prior, respectively.}
\end{table}

\begin{figure}[tbph!]
\centering
\includegraphics[width=0.49\columnwidth]{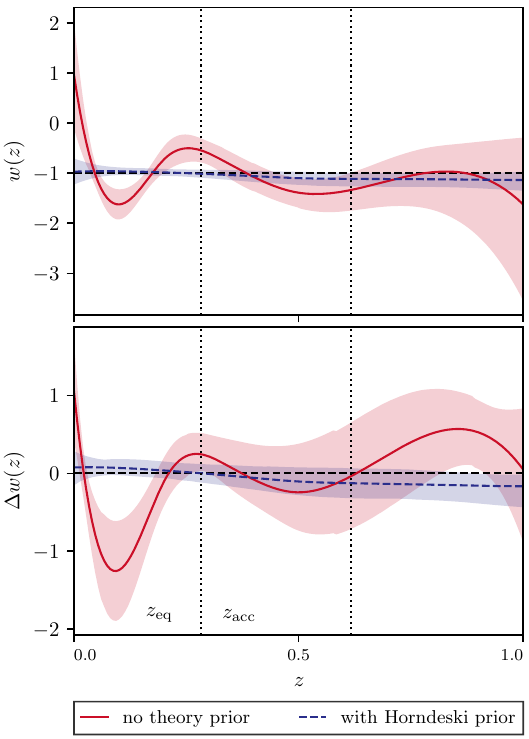}
\includegraphics[width=0.49\columnwidth]{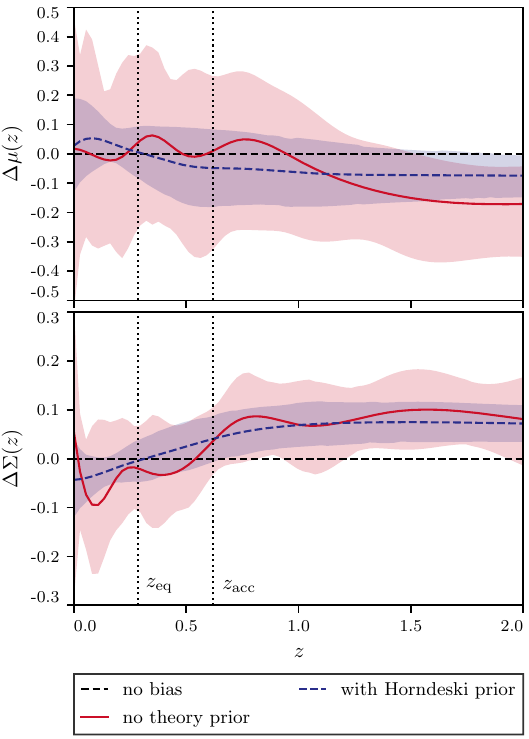}
\caption{\label{fig:ode-mu-sigma} 
{\it Left panel:} the reconstructed equation of state, $w(z)$ (top panel) and the difference between the reconstructed $w(z)$ and the best fit CPL parametrization (bottom panel) with and without using the Horndeski prior. Only $z\in[0,1]$ are shown due to the presence of singularities at higher $z$ caused by negative values of the effective DE density.
{\it Right panel:} the difference between our reconstruction of $\mu$ and $\Sigma$ from Baseline+LSS with and without the Horndeski prior and the corresponding best fit DE fraction parametrization used by DES and Planck. The corresponding Bias values are given in Table.~\ref{tab:bias}.
}
\end{figure}

As mentioned earlier, the projection of $\Omega_X(a)$ onto $w(a)$ is not always well-defined, as the effective DE density can be negative in MG theories and in particular MCMC samples in our reconstruction. In fact, we found that such occurrences were too frequent at $z>1$ in reconstructions without the Horndeski prior, and at $z>2$ when using the prior. Thus, in the case of $w$, we restrict our projections to $z\in [0,1]$ and $[0,2]$, respectively.

To quantify the bias introduced by simple parameterizations, we compute the ``Bias'', defined as
\begin{align} \label{eq:bias}
{\rm Bias} = {\bf d}^T {\cal C}^{-1}_p {\bf d}
\end{align}
where ${\bf d}$ is the vector of differences between the reconstructed values of the nodes of a given function, and the values given by the best fit simple parameterization. 
Bias is bounded from above by the square of the SNR. 
The closer the bias is to its upper bound, the less adequate is the parametrization. Table~\ref{tab:bias} shows the two values for the parameterizations listed above. As one can see, both the constant $w$, and the CPL forms are a very poor representation of DE evolution if no theory prior is used. However, with the Horndeski prior, they perform quite well, as $\Omega_X(a)$ in this case was quite consistent with a constant.

In the case of $\mu$ and $\Sigma$, most parameterizations perform quite poorly, with the exception of the linear model of $\mu(a)$. The DE fraction parametrization, which is one of the most popular ones in the literature, performs the worst. To visualize the bias, we plot the differences between the reconstructed $\mu$ and $\Sigma$ and the best fit DE fraction parameterizations in Fig.~\ref{fig:ode-mu-sigma}. Fig.~\ref{fig:ode-mu-sigma} also shows the difference plot for $w(z)$ in the case of the CPL parametrization.

\section{Summary} \label{Sec:Summary}

In this paper, we examined the joint reconstruction of three time-dependent functions that capture departures from the standard cosmological model, $\Lambda$CDM, at the level of the expansion history as well as gravitational effects on matter and light on linear scales. The background expansion is described in terms of the effective dark energy fractional energy density $\Omega_X(z)$, while gravitation effects in the large scale structure are described by $\mu(z)$ and $\Sigma(z)$, which parametrise the relation of the Newton potential and the lensing potential to the density contrast, respectively. The reconstruction was performed with and without a theoretical prior derived from the Horndeski theory, both for a Baseline dataset (Planck, BAO and SN) and an extended one including DES and RSD. 
Without the Horndeski prior, the reconstruction is affected by an implicit prior imposed by the redshift binning and is prone to high frequency oscillations that are poorly constrained by the data. The theoretical prior successfully suppressed oscillations in the reconstructed functions. All reconstructed functions are consistent with the GR predictions within 2-3$\sigma$. 

Using the Covariant Principal Component (CPC) analysis of the prior and posterior covariances, we determined the number of modes that are constrained by the data relative to the Horndeski prior. Function $\mu$ is the least constrained, with only 2 such modes, while 8 and 6 modes are constrained for $\Sigma$ and $\Omega_X$, respectively. We examined the redshift dependence of the eigenmodes, identifying the features in the data that determines them, and quantified their contributions to deviations from $\Lambda$CDM. When looking for any departure from $\Lambda$CDM, without identifying the function responsible for it, we find that data can constrains 15 {\it combined} eigenmodes of $\mu$, $\Sigma$ and $\Omega_X$.

Overall, our work shows that current data allow for an informative, data-driven, reconstruction of the cosmological model, allowing us to constrain much more than a few parameters of the {\it ad hoc} simple parametrizations that, as we have demonstrated, would lead to significantly biased results.

\acknowledgments 
We thank Jian Li, Simone Peirone and Alex Zucca for their contributions to developing some of the numerical codes used in this work. 
MR is supported in part by NASA ATP Grant No. NNH17ZDA001N and by funds provided by the Center for Particle Cosmology.
LP is supported by the National Sciences and Engineering Research Council (NSERC) of Canada, and by the Chinese Academy of Sciences President's International Fellowship Initiative, Grant No. 2020VMA0020. KK was supported by the European Research Council under the European Union's Horizon 2020 programme (grant agreement No.646702 ``CosTesGrav"). KK is supported the UK STFC grant ST/S000550/1 and ST/W001225/1. MM has received the support of a fellowship from `la Caixa' Foundation (ID 100010434), with fellowship code LCF/BQ/PI19/11690015, the support of the Spanish Agencia Estatal de Investigacion through the grant `IFT Centro de Excelencia Severo Ochoa SEV-2016-0599' and funding by the Agenzia Spaziale Italiana (ASI) under agreement n. 2018-23-HH.0. AS acknowledges support from the NWO and the Dutch Ministry of Education, Culture and Science (OCW). GBZ is supported by the National Key Basic Research and Development Program of China (No. 2018YFA0404503), NSFC Grants 11925303, 11720101004, and a grant of CAS Interdisciplinary Innovation Team.
This research was enabled in part by support provided by WestGrid ({\tt www.westgrid.ca}), Compute Canada ({\tt www.computecanada.ca}) and by the University of Chicago Research Computing Center through the Kavli Institute for Cosmological Physics at the University of Chicago.

\bibliographystyle{JHEP}
\bibliography{mgrecon}

\end{document}